\documentclass[12pt]{article}
  \usepackage{amsthm,amsfonts}
  \usepackage{amsmath}
\newcommand{\bea}   {\begin{eqnarray}}
\newcommand{\eea}   {\end{eqnarray}}

\begin{document}
\renewcommand{\thefootnote}{\fnsymbol{footnote}}

\thispagestyle{empty}
\title{Supersymmetric Extension of Hopf Maps: \\
${\cal N}=4$ $\sigma$-models and the ${\bf S}^3\rightarrow {\bf S}^2$ Fibration.}

\author{L. Faria Carvalho\thanks{{\em e-mail: leofc@cbpf.br}}, Z. Kuznetsova\thanks{{\em e-mail: zhanna.kuznetsova@ufabc.edu.br}}
~and F.
Toppan\thanks{{\em e-mail: toppan@cbpf.br}}
\\
\\
}
\maketitle

\centerline{$^{\ast\ddag}${\it CBPF, Rua Dr. Xavier Sigaud 150, Urca,}}{\centerline {\it\quad
cep 22290-180, Rio de Janeiro (RJ), Brazil.}}
{\centerline{
$^{\dag}${\it UFABC, Rua Catequese 242, Bairro Jardim,}}{\centerline{\it\quad cep 09090-400, Santo Andr\'e (SP), Brazil.}}}

\begin{abstract}
We discuss four off-shell ${\cal N}=4$ $D=1$ supersymmetry transformations, their associated one-dimensional $\sigma$-models and their mutual relations. They are given by \par
{\em I}) the $(4,4)_{lin}$ linear ``root" supermultiplet (supersymmetric extension of ${\bf R}^4$),\par
{\em II}) the $(3,4,1)_{lin}$ linear supermultiplet (supersymmetric extension of ${\bf R}^3$),\par
{\em III}) the $(3,4,1)_{nl}$ non-linear supermultiplet living on ${\bf S}^3$ and\par
{\em  IV}) the $(2,4,2)_{nl}$ non-linear supermultiplet living on ${\bf S}^2$.\par
The $I\rightarrow II$ map is the supersymmetric extension of the ${\bf R}^4\rightarrow {\bf R}^3$ bilinear map, while
the $II\rightarrow IV$ map is the supersymmetric extension of the ${\bf S}^3\rightarrow {\bf S}^2$ first Hopf fibration.
The restrictions on the ${\bf S}^3$, ${\bf S}^2$ spheres are expressed in terms of the stereographic projections. The non-linear supermultiplets, whose supertransformations are local differential polynomials,
are not equivalent to the linear supermultiplets with the same field content.\par
The $\sigma$-models are determined in terms of an unconstrained prepotential of the target coordinates. The
Uniformization Problem requires solving an inverse problem for the prepotential. \par
The basic features of the supersymmetric extension of the second and third Hopf maps are briefly sketched.\par
Finally, the Schur's lemma (i.e. the real, complex or quaternionic property) is extended to all minimal linear supermultiplets up to ${\cal N}\leq8$.
\end{abstract}
\vskip1cm
\hfill{CBPF-NF-011/09}

\newpage
\section{Introduction.}

In recent years the linear off-shell representations of the one-dimensional ${\cal N}$-extended superalgebra underlying the Supersymmetric Quantum Mechanics \cite{wit} have been substantially elucidated \cite{gr}--\cite{dfghilm2}.
The superalgebra admits ${\cal N}$ odd generators $Q_I$ (the supercharges, with $I=1,2,\ldots, {\cal N}$) and a single even generator
(the Hamiltonian $H$), satisfying
\bea\label{superalgebra}
\{Q_I,Q_J\}&=&\delta_{IJ}H,\nonumber\\
\relax [H,Q_I]&=&0.
\eea
A lot of information is now available concerning the minimal linear representations (also known as irreducible representations) \cite{pt}--\cite{kt2} based on the minimal number of time-dependent bosonic (and an equal number of time-dependent fermionic) fields belonging to a (\ref{superalgebra}) supermultiplet, as well as the non-minimal
linear representations (admitting a reducible, but indecomposable set of fields), see \cite{krt,dfghilm,dfghilm2}.\par
The most interesting linear representations can be encoded in graphs \cite{fg,dfghil,dfghil2,kt,kt2} (the fields are visualized by dots, the supertransformations by edges) and are further characterized by their field content (the mass-dimension of the fields entering the graph), their connectivity (number of edges entering dots at a given mass-dimension), etc.\par
A systematic construction of one-dimensional ${\cal N}$-extended supersymmetrically invariant $\sigma$-models is made possible by the knowledge of the linear representations,  see \cite{krt, grt}. This systematic approach, alternative to the standard construction based on constrained superfields (see \cite{bikl, ils}), already proved its usefulness for building ${\cal N}>4$ invariant actions. \par
A large part of this activity is inspired by the so-called ``oxidation program'' \cite{glpr,top}, i.e. the construction of one-dimensional, off-shell supersymmetric actions, invariant for a sufficiently large value of ${\cal N}$, as a preliminary step to construct ``oxidized'' supersymmetric models in higher dimensions. In this context ``off-shell'' is the keyword. A properly conducted oxidation program would require the knowledge of all off-shell realizations of (\ref{superalgebra}). Besides the linear representations, a much less understood class of off-shell realizations of (\ref{superalgebra}) is given by the non-linear realizations (the supertransformations are non-linear functions of the fields entering the supermultiplets and their time-derivatives). Non-linear off-shell realizations have been constructed in the literature with a large variety of methods \cite{il,ikl,bbks,bkm,iva,freedom,root} (using, in most of the cases, the manipulation of superfields). The interested reader can consult the cited papers to appreciate the variety of methods which have been used.

Despite the richness of the results so far obtained (or, better, due to this richness), the status of the non-linear off-shell realizations of (\ref{superalgebra}) is at present somehow confusing. This situation has to be compared with the rather clear picture concerning the linear representations. Some very natural questions can be addressed. One for all: under which condition a non-linear off-shell realization is genuinely non-linear and not a linear representation ``in disguise''? A partial answer to this one and similarly related questions will be given in the following. \par

In this paper we construct genuine non-linear off-shell realizations of the (\ref{superalgebra}) superalgebra
in terms of a very precise geometrical setting, based on the supersymmetric extensions of the Hopf maps.
Essentially, a bilinear map between Euclidean spaces (${\bf R}^{2k}\rightarrow {\bf R}^{k+1}$, for $k=1,2,4,8$) can be supersymmetrically extended to a map connecting one-dimensional, off-shell, linear supermultiplets. The restrictions of these supermultiplets on spheres induce non-linear, off-shell realizations of the supersymmetry (non-linear supermultiplets). The spheres can be parametrized using stereographic projection, hyperspherical coordinates, etc. (in the following we will make use of the stereographic projection which induces non-linear, {\em local}, supertransformations). The supersymmetric extension of the Hopf map is the generalization of the ${\bf S}^{2k-1}\rightarrow {\bf S}^k$ Hopf fibration as the map connecting their associated non-linear off-shell supermultiplets.  An explicit construction is carried out for $k=2$ (corresponding to the first Hopf map) producing ${\cal N}=4$ non-linear off-shell supermultiplets.

The main virtue of this construction is its very natural geometrical setup. On the other hand, nothing can be said concerning other possible non-linear off-shell supermultiplets not arising from the supersymmetric extension of a Hopf fibration. It is quite likely that the $(3,4,1)_{nl}$ and the $(2,4,2)_{nl}$ ${\cal N}=4$ non-linear off-shell supermultiplets introduced in the following are related with the non-linear supermultiplets of the same field content already appearing in the literature \cite{root, bny}
and also constructed in association with a Hopf map, but with a more pragmatic approach. This possible equivalence would require an investigation of its own and will not be conducted here. Instead, in this paper we further use the geometrical setup to construct one-dimensional ${\cal N}=4$-invariant $\sigma$ models based on the non-linear supertransformations and investigate the Uniformization problem associated to their bosonic limit. The present scheme can be in principle further applied to the supersymmetric extensions of the second and third Hopf maps. Some comments on that are made in the text.

The bosonic Hopf fibrations are based on maps which are $U(1)$-invariant (for $k=2$) or $SU(2)$-invariant
(for $k=4$). This invariance is closely related with the Schur's lemma applied to Clifford algebras \cite{oku}. The Schur's lemma admits a natural extension to minimal, linear off-shell supermultiplets.
For completeness, we present here for the first time the supersymmetric extension of the Schur's lemma to all minimal linear supermultiplets up to ${\cal N}\leq 8$.

We should mention that in a different context and framework (in application to higher-dimensional quantum Hall systems \cite{zh}) supersymmetric extensions of Hopf maps were considered in \cite{has1}, while in \cite{has2}
non-compact manifolds, associated to split-algebras, were investigated. In \cite{bbl} a superextension of the Dirac monopole was obtained in terms of superfiber bundle. 
Several works, see e.g. \cite{bbkno,kno,bko,filech,kl,bks} and references therein, have discussed physical applications of supersymmetric systems and their relation with supermultiplets.

We postpone to the Conclusions a more detailed discussion of the results here found and of the lines of research that they open.

The scheme of the paper is as follows. In the next Section the bosonic Hopf fibrations will be reviewed on the basis of the formalism which in the following will be supersymmetrically extended. In Section {\bf 3} linear and
non-linear ${\cal N}=4$ off-shell supermultiplets are induced from the first Hopf fibration. In Section {\bf 4} the mappings connecting these ${\cal N}=4$ off-shell realizations are explicitly presented. We discuss in Section {\bf 5} the construction of the ${\cal N}=4$ off-shell supersymmetric invariant actions expressed, for each off-shell supermultiplet, in terms of an unconstrained prepotential.
The connection of the invariant actions with the one-dimensional $\sigma$-models arising as bosonic limits
and the Uniformization problem are explained in Section {\bf 6}. In Section {\bf 7} some features of the supersymmetric extensions of the second and third Hopf maps are outlined. In Section {\bf 8} we mention some possible applications of the construction of one-dimensional off-shell supersymmetry realizations for the oxidation program, namely the reconstruction of higher-dimensional supersymmetric theories from one-dimensional data.
For completeness, in Section {\bf 9} we extend the Schur's lemma to all off-shell, minimal, linear supermultiplets up to ${\cal N}=8$. In the Conclusions we make some comments on our construction and outline some further possible applications. In the Appendix we present, for completeness, the ${\cal N}=4$ non-linear supermultiplet $(3,4,1)_{nl}$ in a quaternionic covariant form with one supercharge linearly realized.

\section{Hopf fibrations. The bosonic case.}

The four Hopf maps (for $k=1,2,4,8$) can be illustrated by the following commutative diagram
\bea\label{hopfd1}
&
\begin{array}{lll}
 {\bf R}^{2k}&
 \stackrel{p}{\longrightarrow}&{\bf R}^{k+1}\\
~^{_{\rho}}{\downarrow}&&\downarrow^{_{\rho'}}\\
{\bf S}^{2k-1}&\stackrel{h}{\longrightarrow}&{\bf S}^{k}
\end{array}&
\eea
which connects four spaces (two Euclidean spaces and two spheres) which, for later convenience, can be identified as $I,II,III,IV$ according to
\bea\label{hopfd2}
&
\begin{array}{ccc}
 I&
 \stackrel{p}{\longrightarrow}&II\\
 ~^{_{\rho}}\downarrow\quad&&\downarrow^{_{\rho'}}\\
III&\stackrel{h}{\longrightarrow}&IV
\end{array}&
\eea
The four arrows correspond to the following maps:\\
- the bilinear map $p: I\rightarrow II$, sending coordinates ${\vec u}\in {\bf R}^{2k}$ into coordinates
${\vec x}\in {\bf R}^{k+1}$ according to
\bea\label{bilin}
p: && {\vec u} \mapsto {x_i} = u^T\gamma_iu,
\eea
($\gamma_i$ are the Euclidean gamma matrices of ${\bf R}^{k+1}$);\\
- the restrictions $\rho,\rho'$ on spheres, where $\rho: I\rightarrow III$ and $\rho': II\rightarrow IV$;\\
- the hopf map $h: II\rightarrow IV$, admitting ${\bf S}^{k-1}$ as a fiber (for $k=8$, ${\bf S}^7$ is a parallelizable manifold but not, properly speaking, a group-manifold due to the nonassociativity of the octonions; for $k=1$, ${\bf S}^0\equiv {\bf Z}_2$).\par
For $k=1,2,4,8$ the map (\ref{bilin}) preserves the norm, allowing to induce the map $h$ from $p$:
\bea\label{spheres}
u^Tu=R\mapsto x^Tx=r, &{with}& r=R^2.
\eea
By setting $k=2^l$, the four Hopf maps $h$ will be referred to (for $l=0,1,2,3$ respectively) as the $0^{th}, 1^{st}, 2^{nd}$ and $3^{rd}$ Hopf map. \par
In the following we will give a detailed description of the supersymmetric extension of the first Hopf map ($k=2$), corresponding to the diagram
\bea
&
\begin{array}{cll}\label{hopf1}
 {\bf R}^{4}&
 \stackrel{p}{\longrightarrow}\quad \quad&{\bf R}^{3}\\
 ~^{_{\rho}}\downarrow&&\downarrow^{_{\rho'}}\\
{\bf S}^{3}&\stackrel{h}{\longrightarrow}&{\bf S}^{2}
\end{array}&
\eea
In the supersymmetric extension ${\bf R}^4$ is replaced by the ${\cal N}=4$ root supermultiplet $(4,4)$ whose four bosonic (target) coordinates correspond to the coordinates of ${\bf R}^4$. The off-shell supermultiplets extending $II$, $III$ and $IV$ are induced by applying, respectively, the map $p$ and the restriction $\rho$ to $(4,4)$, as well as the restriction $\rho'$ on the induced supermultiplet generalizing $III$. \par
For our purposes it will be convenient to define the target coordinates of the supermultiplets extending $III$ ($IV$) in terms of the stereographic projection of the $I$ ($II$) target coordinates.\par
For $k=2$ we can express the three Euclidean gamma matrices $\gamma_i$ as
\bea\label{gamma}
&\gamma_1=\tau_1\otimes {\bf 1}_2, \quad \gamma_2=\tau_A\otimes\tau_A,\quad\gamma_3=\tau_2\otimes{\bf 1}_2,&
\eea
where
\begin{eqnarray}\label{tau}
\tau_1 = \left(\begin{array}{cc}
0 & 1 \\
1 & 0 \\
\end{array}
\right),\quad
\tau_2 = \left(
\begin{array}{cc}
1 & 0 \\
0 & -1 \\
\end{array}
\right),\quad
\tau_A = \left(
\begin{array}{cc}
0 & 1 \\
-1 & 0 \\
\end{array}
\right).&
\end{eqnarray}
With this convention the bilinear map $p$ is explicitly presented as
\bea\label{bilin2}
x_1&=&2(u_1u_3+u_2u_4),\nonumber\\
x_2&=&2(u_1u_4-u_2u_3),\nonumber\\
x_3&=& {u_1}^2+{u_2}^2-{u_3}^2-{u_4}^2.
\eea
It is invariant under the $\sigma$ transformation $(\sigma^2=-{\bf 1}$), given by
\bea\label{u1}
\sigma:&u_1\mapsto u_2,\quad u_2\mapsto -u_1,\quad u_3\mapsto u_4,\quad u_4\mapsto -u_3.&
\eea
\section{Linear and non-linear off-shell supermultiplets.}

The construction of the supersymmetric extension of the $1^{st}$ Hopf map has been outlined in the previous Section. Here we present and discuss the results.\par
Induced by the ${\cal N}=4$ $(4,4)$ root supermultiplet, three more (inequivalent) ${\cal N}=4$ off-shell supermultiplets are obtained. They correspond to the supersymmetric extensions of $II$, $III$ and $IV$.
The supertransformations extending $II$ are all linear. This supermultiplet has field content $(3,4,1)$ and will therefore be denoted as $(3,4,1)_{lin}$. The supermultiplets extending $III$ and $IV$ possess supertransformations which are differential polynomials in their component fields. Since their field content is, respectively, $(3,4,1)$ and $(2,4,2)$, the supermultiplets will be denoted as $(3,4,1)_{nl}$ and $(2,4,2)_{nl}$, respectively.\par
Schematically, we have
  \bea\label{n4mlt}
&
\begin{array}{ccc}
 (4,4)&
 \longrightarrow&(3,4,1)_{lin}\\
 \downarrow&&\downarrow\\
(3,4,1)_{nl}&\longrightarrow&(2,4,2)_{nl}
\end{array}&
\eea
Their component fields are parametrized according to
\bea\label{susyext1}
&
\begin{array}{ccc}
 \begin{array}{c}
 (u_1,u_2,u_3,u_4;\psi_1,\psi_2,\psi_3,\psi_4)
 \end{array}&
 \longrightarrow&\begin{array}{c}
 (x_1,x_2,x_3;\mu_1,\mu_2,\mu_3,\mu_4; f)
 \end{array}\\
 \downarrow&&\downarrow\\
\begin{array}{c}
 (w_1,w_2,w_3;\xi_1,\xi_2,\xi_3,\xi_4;g)
 \end{array}&\longrightarrow&\begin{array}{c}
 (z_1,z_2;\eta_1,\eta_2,\eta_3,\eta_4;h_1,h_2)
 \end{array}
\end{array}&
\eea
The greek letters have been employed to denote the fermionic fields; ${\vec u}, {\vec x}, {\vec w}, {\vec z}$ denote the bosonic target coordinates of the respective supermultiplets, while $f,g,h_{1,2}$ denote the auxiliary fields.\par
We present at first the supersymmetry transformations of the four supermultiplets above (the presentation of the transformations explicitly connecting their component fields, namely the ``arrows'' in (\ref{susyext1}), will be given in the next Section). In the following tables the entries give the supertransformations of the component fields under the action of the $Q_I$ supersymmetry operator.\par
The ${\cal N}=4$ (linear) $(4,4)$ root supermultiplet can be explicitly presented as
\bea\label{table1}
&
\begin{array}{|c|c|c|c|c|}\hline
&Q_1&Q_2&Q_3&Q_4\\ \hline
u_1&\psi_1&\psi_2&\psi_3&\psi_4\\
u_2&\psi_2&-\psi_1&\psi_4&-\psi_3\\
u_3&\psi_3&-\psi_4&-\psi_1&\psi_2\\
u_4&\psi_4&\psi_3&-\psi_2&-\psi_1\\ \hline
\psi_1&\dot{u}_1&-\dot{u}_2&-\dot{u}_3&-\dot{u}_4\\
\psi_2&\dot{u}_2&\dot{u}_1&-\dot{u}_4&\dot{u}_3\\
\psi_3&\dot{u}_3&\dot{u}_4&\dot{u}_1&-\dot{u}_2\\
\psi_4&\dot{u}_4&-\dot{u}_3&\dot{u}_2&\dot{u}_1\\ \hline
\end{array}
&
\eea
The $(3,4,1)_{lin}$ supermultiplet is given by
\bea\label{table2}
&
\begin{array}{|c|c|c|c|c|}\hline
&Q_1&Q_2&Q_3&Q_4\\ \hline
x_1&\mu_1&-\mu_2&-\mu_3&\mu_4\\
x_2&\mu_2&\mu_1&-\mu_4&-\mu_3\\
x_3&\mu_3&\mu_4&\mu_1&\mu_2\\ \hline
\mu_1&\dot{x}_1&\dot{x}_2&\dot{x}_3&-f\\
\mu_2&\dot{x}_2&-\dot{x}_1&f&\dot{x}_3\\
\mu_3&\dot{x}_3&-f&-\dot{x}_1&-\dot{x}_2\\
\mu_4&f        &\dot{x}_3&-\dot{x}_2&\dot{x}_1\\ \hline
f&\dot{\mu}_4  &-\dot{\mu}_3&\dot{\mu}_2&-\dot{\mu}_1\\ \hline
\end{array}
&
\eea
The $(3,4,1)_{nl}$ supermultiplet is given by

\bea\label{table3}
&
\begin{array}{|c|c|c|c|c|}\hline
&Q_1&Q_2&Q_3&Q_4\\ \hline
w_1&\xi_1+\frac{1}{R}(w_1\xi_4)&\xi_2+\frac{1}{R}(w_1\xi_3)&\xi_2-\frac{1}{R}(w_1\xi_2)&\xi_4-\frac{1}{R}(w_1\xi_1)\\
w_2&\xi_2+\frac{1}{R}(w_2\xi_4)&-\xi_1+\frac{1}{R}(w_2\xi_3)&\xi_4-\frac{1}{R}(w_2\xi_2)&-\xi_3-\frac{1}{R}(w_2\xi_1)\\
w_3&\xi_3+\frac{1}{R}(w_3\xi_4)&-\xi_4+\frac{1}{R}(w_3\xi_3)&-\xi_1-\frac{1}{R}(w_3\xi_2)&\xi_2-\frac{1}{R}(w_3\xi_1)\\ \hline
\xi_1
&\dot{w}_1-\frac{1}{R}(w_1g+\xi_1\xi_4)   &-\dot{w}_2+\frac{1}{R}(w_2g-\xi_1\xi_3) &-\dot{w}_3+\frac{1}{R}(w_3g+\xi_1\xi_2) &-g\\
\xi_2
&\dot{w}_2-\frac{1}{R}(w_2g+\xi_2\xi_4) &\dot{w}_1-\frac{1}{R}(w_1g+\xi_2\xi_3) &-g&\dot{w}_3-\frac{1}{R}(w_3g+\xi_1\xi_2) \\
\xi_3
&\dot{w}_3-\frac{1}{R}(w_3g+\xi_3\xi_4) &g
&\dot{w}_1-\frac{1}{R}(w_1g+\xi_2\xi_3) &-\dot{w}_2+\frac{1}{R}(w_2g-\xi_1\xi_3) \\
\xi_4
&g      &-\dot{w}_3+\frac{1}{R}(w_3g+\xi_3\xi_4)
&\dot{w}_2-\frac{1}{R}(w_2g+\xi_2\xi_4) &\dot{w}_1-\frac{1}{R}(w_1g+\xi_1\xi_4) \\ \hline
g&\dot{\xi}_4  &-\dot{\xi}_3&-\dot{\xi}_2&-\dot{\xi}_1\\ \hline
\end{array}
&\nonumber\\
&&
\eea
Finally, the $(2,4,2)_{nl}$ supermultiplet is given by
\bea\label{table4}
&
\begin{array}{|c|c|c|c|c|}\hline
&Q_1&Q_2&Q_3&Q_4\\ \hline
z_1&\eta_1+\frac{1}{r}(z_1\eta_3)&-\eta_2+\frac{1}{r}(z_1\eta_4)&-\eta_3+\frac{1}{r}(z_1\eta_1)&\eta_4+\frac{1}{r}(z_1\eta_2)\\
z_2&\eta_2+\frac{1}{r}(z_2\eta_3)&\eta_1+\frac{1}{r}(z_2\eta_4)&-\eta_4+\frac{1}{r}(z_2\eta_1)&-\eta_3+\frac{1}{r}(z_2\eta_2)\\ \hline
\eta_1
&\dot{z}_1-\frac{1}{r}(z_1h_1+\eta_1\eta_3)   &\dot{z}_2-\frac{1}{r}(z_2h_1+\eta_1\eta_4) &h_1&-h_2-\frac{1}{r}(\eta_1\eta_2) \\
\eta_2
&\dot{z}_2-\frac{1}{r}(z_2h_1+\eta_2\eta_3) &-\dot{z}_1+\frac{1}{r}(z_1h_1-\eta_2\eta_4) &h_2-\frac{1}{r}(\eta_1\eta_2)&h_1 \\
\eta_3
&h_1 &-h_2-\frac{1}{r}(\eta_3\eta_4)
&-\dot{z}_1+\frac{1}{r}(z_1h_1+\eta_1\eta_3) &-\dot{z}_2+\frac{1}{r}(z_2h_1+\eta_2\eta_3) \\
\eta_4
&h_2-\frac{1}{r}(\eta_3\eta_4)    &h_1
&-\dot{z}_2+\frac{1}{r}(z_2h_1+\eta_1\eta_4) &\dot{z}_1-\frac{1}{r}(z_1h_1+\eta_2\eta_4) \\ \hline
h_1&\dot{\eta}_3  &\dot{\eta}_4&\dot{\eta}_1&\dot{\eta}_2\\
h_2&\dot{\eta}_4 -\frac{1}{r}(h_1\eta_4-h_2\eta_3) &-\dot{\eta}_3+\frac{1}{r}(h_1\eta_3+h_2\eta_4)&\dot{\eta}_2-\frac{1}{r}(h_1\eta_2+h_2\eta_1)&-\dot{\eta}_1+\frac{1}{r}(h_1\eta_1+h_2\eta_2)\\ \hline
\end{array}
&\nonumber\\
&&
\eea
A few comments are in order:\\
- The non-linearity of the $(3,4,1)_{nl}$ and $(2,4,2)_{nl}$ supertransformations
is the mildest possible nonlinearity, since at most bilinear combinations of the component fields appear in the entries.\\
- The constant parameters $R$ (entering $(3,4,1)_{nl}$) and $r$ (entering $(2,4,2)_{nl}$) can be reabsorbed
(set equal to $1$) through a suitable rescaling of the component fields. It is however convenient to present them explicitly to show that in the contraction limit (for $R,r\rightarrow \infty$) the linear supermultiplets $(3,4,1)_{lin}$ and, respectively, $(2,4,2)_{lin}$ are recovered. As a consequence,
$(3,4,1)_{nl}$ and $(2,4,2)_{nl}$ are more general than the corresponding linear supermultiplets with the same field content. Indeed, while the latter can be recovered from the non-linear ones, the converse is not true, as it will be clear from the discussion at the end of the next Section.
\par
We can summarize the results of this Section as follows. Besides the ${\cal N}=4$ minimal linear supermultiplets classified in \cite{pt}, at least two extra, non-linear, inequivalent supermultiplets are found. They are geometrically induced by the restriction of the $(4,4)$ and $(3,4,1)_{lin}$ linear supermultiplets on spheres. The inequivalent supermultiplets of field content $(3,4,1)$ ($(2,4,2)$) are determined in terms of a parameter $\epsilon=0,1$ (basically, $\epsilon=\frac{1}{R}$ or, respectively $\epsilon=\frac{1}{r}$, with the radius of the sphere suitably normalized, either $R=1$ or $R=\infty$).
We can write $(3,4,1)_{lin}\equiv (3,4,1)_{\epsilon=0}$, $(3,4,1)_{nl}\equiv (3,4,1)_{\epsilon=1}$ and analogous relations for the $(2,4,2)$ field content.

\section{The ``non-linear dressing".}

We present here the transformations connecting the supermultiplets introduced in the previous Section
(the ``arrows'' in the (\ref{susyext1}) diagram).\par
The map $(4,4)\rightarrow (3,4,1)_{lin}$ is the supersymmetric extension of the bilinear bosonic transformation $p$ given in (\ref{bilin}). $(4,4)\rightarrow (3,4,1)_{lin}$ is given by
\bea\label{susybilin}
x_1&=&2(u_1u_3+u_2u_4),\nonumber\\
x_2&=&2(u_1u_4-u_2u_3),\nonumber\\
x_3&=& {u_1}^2+{u_2}^2-{u_3}^2-{u_4}^2,\nonumber\\
\mu_1&=&2(u_1\psi_3+u_2\psi_4+u_3\psi_1+u_4\psi_2),\nonumber\\
\mu_2&=& 2(u_1\psi_4-u_2\psi_3-u_3\psi_2+u_4\psi_1),\nonumber\\
\mu_3&=&2(u_1\psi_1+u_2\psi_2-u_3\psi_3-u_4\psi_4),\nonumber\\
\mu_4&=& 2(u_1\psi_2-u_2\psi_1+u_3\psi_4-u_4\psi_3),\nonumber\\
f&=&2(u_1{\dot u}_2-u_2{\dot u}_1+u_3{\dot u}_4-u_4{\dot u}_3)+4(\psi_1\psi_2+\psi_3\psi_4).
\eea
We should stress the fact that the $(3,4,1)_{lin}$ supermultiplet is recovered from $(4,4)$ also via another
transformation, the linear dressing discussed in \cite{pt} which, essentially, allows to identify the auxiliary field $f$ with a time-derivative of one of the $u_i$'s fields; let's say $f={\dot u}_4$.
We have therefore a double derivation of ${\cal N}=4$ $(3,4,1)_{lin}$ from the ${\cal N}=4$ $(4,4)$ root supermultiplet.\par
The transformation $(4,4)\rightarrow (3,4,1)_{nl}$ is induced after identifying the three target coordinates
entering $(3,4,1)_{nl}$ with the coordinates of the stereographic projection of the ${\bf S}^3$ sphere embedded in ${\bf R}^4$. For $(4,4)\rightarrow (3,4,1)_{nl}$ we obtain, explicitly,
\bea\label{susy13}
w_i&=&\frac{Ru_i}{R-u_4},\quad for\quad i=1,2,3,\nonumber\\
\xi_j&=&\frac{R\psi_j}{R-u_4},\quad for \quad j=1,2,3,4\nonumber\\
g&=& \frac{R{\dot u}_4}{R-u_4}.
\eea
Similarly, the transformation $(3,4,1)_{lin}\rightarrow (2,4,2)_{nl}$ is induced after identifying the
two target coordinates entering $(2,4,2)_{nl}$ with the coordinates of the stereographic projection of the
${\bf S}^2$ sphere embedded in ${\bf R}^3$. For $(3,4,1)_{lin}\rightarrow (2,4,2)_{nl}$ we have
\bea\label{susy24}
z_i&=&\frac{rx_i}{r-x_3},\quad for\quad i=1,2,\nonumber\\
\eta_j&=&\frac{r\mu_j}{r-x_3},\quad for \quad j=1,2,3,4,\nonumber\\
h_1&=& \frac{r{\dot x}_3}{r-x_3},\nonumber\\
h_2&=&\frac{rf}{r-x_3}.
\eea
The last transformation connects the component fields of the two nonlinear supermultiplets. It corresponds to
a nonlinear version of the dressing transformation.
For $(3,4,1)_{nl}\rightarrow (2,4,2)_{nl}$ we have
\bea\label{susy34}
z_i&=&\frac{r}{R}w_i,\quad for \quad i=1,2,\nonumber\\
\eta_j&=&\frac{r}{R}\xi_j,\quad for\quad j=1,2,3,4,\nonumber\\
h_1&=& \frac{r}{R}g,\nonumber\\
h_2&=&\frac{r}{R}({\dot w}_3-w_3g).
\eea

This transformation requires the identifications
${Q_1}^{IV}={Q_1}^{III}$ and ${Q_l}^{IV}=-{Q_l}^{III}$ for $l=2,3,4$.\par
Some comments are in order. The (\ref{susy13}) map is not covariant with respect to the $u(1)$ ($\sigma^2=-{\bf 1}$) action given by the combination of (\ref{u1}) on the bosonic component of $(4,4)$, together with
\bea
\sigma:&\psi_1\mapsto \psi_2,\quad \psi_2\mapsto -\psi_1,\quad \psi_3\mapsto \psi_4,\quad \psi_4\mapsto -\psi_3,&
\eea
on the fermionic component fields of $(4,4)$. \par
Imposing a $u(1)$-covariance would have required another parametrization for the spheres. Indeed, taking a hyperspherical parametrization for the bosonic target coordinates we would have obtained, by construction, a $u(1)$-covariant description of the supertransformations of the nonlinear supermultiplets. There is a price to be paid, however. Using the hypersperical parametrization the supertransformations are nonlocal differential functions involving quotients of trigonometrical functions of the component fields. The use of the stereographic projection, as recalled, gives us local supertransformations, expressed as differential polynomials of the component fields. In this paper we put the emphasis on the non-linear realizations of the off-shell supersymmetry, rather than on the ${\cal N}=4$ supersymmetric systems with global $U(1)$ invariance (supersymmetric systems with gauged $U(1)$ were analyzed in \cite{gauging1,gauging2,gauging3}). In this context the stereographic projection is the natural choice due to its much simpler (and local) description of the nonlinear supersymmetry.  The use of the stereographic projection allows to positively answer the question concerning the genuine nonlinearity of $(3,4,1)_{nl}$ and $(2,4,2)_{nl}$. The nonlinearity of their supertransformations is not ``fake''. Stated otherwise, these supermultiplets {\em are not} linear supermultiplets ``in disguise'', due to an awkwardly chosen reparametrization of the component fields of the linear supermultiplets. For both
$(3,4,1)_{nl}$ and $(2,4,2)_{nl}$ the best one can do is to realize linearly, via reparametrization of the component fields, at most one of the four supercharges $Q_I$. The three remaining supercharges are necessarily realized non-linearly. A quaternionic covariant presentation of $(3,4,1)_{nl}$
with one linearly realized supercharge and three non-linear supercharges is explicitly shown in the Appendix.\par
The component fields of $(3,4,1)_{nl}$ and $(2,4,2)_{nl}$ are obtained (just like $(3,4,1)_{lin}$ and $(2,4,2)_{lin}$) as functions of the $(4,4)$ component fields and their time-derivatives. However (this is the crucial feature), they cannot be obtained as functions of the component fields (and their time derivatives) which belong to the linear supermultiplets possessing the same field content. There is no local map sending $(3,4,1)_{lin}\rightarrow (3,4,1)_{nl}$ or viceversa. \par
It is worth recalling here that the field content is an important physical characterization of the supersymmetric system. It gives in particular the dimensionality (number of target coordinates) of the target manifold induced by a one-dimensional supersymmetric invariant $\sigma$-model (we recall that one-dimensional supersymmetric $\sigma$-models were first considered in \cite{sigma1,sigma2}).

\section{Invariant supersymmetric actions.}

We summarize at first the results derived so far. The supersymmetric extension of the $1^{st}$ Hopf map produces four
(two linear and two non-linear) off-shell realizations of the $D=1$ ${\cal N}=4$ supersymmetry. The off-shell realizations, denoted as the supermultiplets $(4,4)$, $(3,4,1)_{lin}$, $(3,4,1)_{nl}$, $(2,4,2)_{nl}$, are interconnected via the transformations obtained in the previous Section.\par
We use now these ingredients to construct four (in association with each one of the above supermultiplets) ${\cal N}=4$-invariant one-dimensional supersymmetric $\sigma$-models whose target manifold is parametrized by the target coordinates of the associated supermultiplet. These $\sigma$-model are determined in terms of an unconstrained function (the prepotential $F$) of the target coordinates. The construction of the invariant action with the correct mass-dimension of the kinetic term (no higher derivatives) follows the prescription given in references \cite{krt} and \cite{grt}. For our purposes here it is sufficient to recall that the supersymmetry operators $Q_I$ act as odd derivatives which satisfy the graded Leibniz rule.\par
The invariant action ${\cal S}=\frac{1}{m}\int dt {\cal L}$ is expressed through the lagrangian ${\cal L}$ s.t.
\bea\label{lagr}
{\cal L} &=& Q_1Q_2Q_3Q_4 (F),
\eea
where $F$ is the unconstrained prepotential.\par
By specializing the above formula in terms of the given off-shell supermultiplets we obtain that, for
$(4,4)$, the corresponding lagrangian ${\cal L}_I$ is given, up to a total derivative, by
\bea\label{lagr1}
{\cal L}_I&=&\Phi(\dot{u_{{1}}}^2+\dot{u_{{2}}}^2+\dot{u_{{3}}}^2+\dot{u_{{4}}}^2-
\psi_{{1}}\dot{\psi_{{1}}}-\psi_{{2}}\dot{\psi_{{2}}}-\psi_{{3}}\dot{\psi_{{3}}}-
\psi_{{4}}\dot{\psi_{{4}}})+\nonumber\\
&&
(\partial_{1}\Phi)(\dot{u_{{2}}}(\psi_{{1}}\psi_{{2}}+\psi_{{3}}\psi_{{4}})
+\dot{u_{{3}}}(\psi_{{1}}\psi_{{3}}-\psi_{{2}}\psi_{{4}})
+\dot{u_{{4}}}(\psi_{{1}}\psi_{{4}}+\psi_{{2}}\psi_{{3}}))+\nonumber\\&&
(\partial_{2}\Phi)(-\dot{u_{{1}}}(\psi_{{1}}\psi_{{2}}+\psi_{{3}}\psi_{{4}})
+\dot{u_{{3}}}(\psi_{{1}}\psi_{{4}}+\psi_{{2}}\psi_{{3}})
-\dot{u_{{4}}}(\psi_{{1}}\psi_{{3}}-\psi_{{2}}\psi_{{4}}))+\nonumber\\&&
(\partial_{3}\Phi)(-\dot{u_{{1}}}(\psi_{{1}}\psi_{{3}}-\psi_{{2}}\psi_{{4}})
-\dot{u_{{2}}}(\psi_{{1}}\psi_{{4}}+\psi_{{2}}\psi_{{3}})
-\dot{u_{{4}}}(\psi_{{1}}\psi_{{2}}+\psi_{{3}}\psi_{{4}}))+\nonumber\\&&
(\partial_{4}\Phi)(-\dot{u_{{1}}}(\psi_{{1}}\psi_{{4}}+\psi_{{2}}\psi_{{3}})
+\dot{u_{{2}}}(\psi_{{1}}\psi_{{3}}-\psi_{{2}}\psi_{{4}})
-\dot{u_{{3}}}(\psi_{{1}}\psi_{{2}}+\psi_{{3}}\psi_{{4}}))+\nonumber\\&&
(\Box \Phi)(\psi_{{1}}\psi_{{2}}\psi_{{3}}\psi_{{4}}),
\eea
where $\Phi$ is determined in terms of the prepotential through
\bea\label{phi1}
\Phi&=&\partial_{1}^{2}F \left( u_{{1}},u_{{2}},u_{{3}},u_{{4}} \right)+\partial_{2}^{2}F \left( u_{{1}},u_{{2}},u_{{3}},u_{{4}} \right)+\partial_{3}^{2}F \left( u_{{1}},u_{{2}},u_{{3}},u_{{4}} \right)+\partial_{4}^{2}F \left( u_{{1}},u_{{2}},u_{{3}},u_{{4}} \right)\nonumber\\&=&\Box F\left( u_{{1}},u_{{2}},u_{{3}},u_{{4}} \right).
\eea
For $(3,4,1)_{lin}$, the corresponding lagrangian ${\cal L}_{II}$ is given by
\bea\label{lagr2}
{\cal L}_{II}&=&\Phi(\dot{x_{{1}}}^2+\dot{x_{{2}}}^2+\dot{x_{{3}}}^2+f^2-
\mu_{{1}}\dot{\mu_{{1}}}-\mu_{{2}}\dot{\mu_{{2}}}-\mu_{{3}}\dot{\mu_{{3}}}-
\mu_{{4}}\dot{\mu_{{4}}})+\nonumber\\
&&
(\partial_{1}\Phi)(\dot{x_{{2}}}(\mu_{{1}}\mu_{{2}}+\mu_{{3}}\mu_{{4}})
+\dot{x_{{3}}}(\mu_{{1}}\mu_{{3}}-\mu_{{2}}\mu_{{4}})
+f(\mu_{{1}}\mu_{{4}}+\mu_{{2}}\mu_{{3}}))+\nonumber\\&&
(\partial_{2}\Phi)(-\dot{x_{{1}}}(\mu_{{1}}\mu_{{2}}+\mu_{{3}}\mu_{{4}})
+\dot{x_{{3}}}(\mu_{{1}}\mu_{{4}}+\mu_{{2}}\mu_{{3}})
-f(\mu_{{1}}\mu_{{3}}-\mu_{{2}}\mu_{{4}}))+\nonumber\\&&
(\partial_{3}\Phi)(-\dot{x_{{1}}}(\mu_{{1}}\mu_{{3}}-\mu_{{2}}\mu_{{4}})-
\dot{x_{{2}}}(\mu_{{1}}\mu_{{4}}+\mu_{{2}}\mu_{{3}})
+f(\mu_{{1}}\mu_{{2}}+\mu_{{3}}\mu_{{4}}))+\nonumber\\&&
(\Box \Phi)(\mu_{{1}}\mu_{{2}}\mu_{{3}}\mu_{{4}}).
\eea
In this case we have the position
\bea\label{phi2}
\Phi &=&\partial_{1}^{2}F \left( x_{{1}},x_{{2}},x_{{3}} \right)+\partial_{2}^{2}F \left( x_{{1}},x_{{2}},x_{{3}} \right)+\partial_{3}^{2}F \left( x_{{1}},x_{{2}},x_{{3}} \right)=\Box F\left( x_{{1}},x_{{2}},x_{{3}} \right).
\eea
For what concerns the non-linear supermultiplets we proceed in the same way. For $(3,4,1)_{nl}$ we construct the invariant action, for simplicity, in terms of an $SO(3)$-invariant prepotential that, without loss of generality, can be expressed ($F(\rho)$) as a function of the coordinate $\rho$ s.t.
\bea\label{rho3}
\rho&=&\sqrt{w_{1}^2+w_{2}^2+w_{3}^2}.
\eea
It is convenient to introduce $A(\rho)$ through
\bea\label{arho}
A(\rho)&=&\frac{d}{d\rho}F(\rho)\equiv F'
\eea
(in the following a prime denotes a $\rho$-derivative). \par
Up to a total derivative, the lagrangian ${\cal L}_{III}$ derived from (\ref{lagr}) is
\bea\label{lagr3}
{\cal L}_{III}&=& \Big[\frac{2}{\rho}A+A'\Big]\big(\dot w_{1} ^2+\dot w_{2} ^2+\dot w_{3} ^2+g^2-\xi_{1}\dot\xi_{1}-\xi_{2}\dot\xi_{2}-
\xi_{3}\dot\xi_{3}-\xi_{4}\dot\xi_{4}\big)+\nonumber\\&&\frac{1}{\rho}\Big[\frac{2}{\rho^2}A-\frac{2}{\rho}A'-A''\Big]
\Big[\dot w_{1}\Big(w_{2}\big(\xi_{1}\xi_{2}+\xi_{3}\xi_{4}\big)+w_{3}\big(\xi_{1}\xi_{3}-\xi_{2}\xi_{4}\big)\Big)-\nonumber\\
&&-\dot w_{2}\Big(w_{1}\big(\xi_{1}\xi_{2}+\xi_{3}\xi_{4}\big)-w_{3}\big(\xi_{1}\xi_{4}+\xi_{2}\xi_{3}\big)\Big)-\dot w_{3}
\Big(w_{1}\big(\xi_{1}\xi_{3}-\xi_{2}\xi_{4}\big)+w_{2}\big(\xi_{1}\xi_{4}+\xi_{2}\xi_{3}\big)\Big)-\nonumber\\&&-g\Big(w_{1}\Big(\xi_{1}\xi_{4}
+\xi_{3}\xi_{4}\Big)-w_{2}\Big(\xi_{1}\xi_{3}-\xi_{2}\xi_{4}\Big)
+w_{3}\big(\xi_{1}\xi_{2}+\xi_{3}\xi_{4}\big)\Big)\Big]+\Big[\frac{4}{\rho}A''+A'''\Big]\xi_{1}\xi_{2}\xi_{3}\xi_{4}+\nonumber\\&&
\frac{1}{R} \bigg\{ -\Big[ \frac{2}{\rho}A+A' \Big]\big(w_{1}\dot w_{1}+w_{2}\dot w_{2}+w_{3}\dot w_{3}\big)g-
\Big[\frac{4}{\rho}A+4A'+\rho A''\Big] \cdot\nonumber \\&& \cdot \Big(\dot w_{1}\big(\xi_{1}\xi_{4}+\xi_{2}\xi_{3}\big)-
\dot w_{2}\big(\xi_{1}\xi_{3}-\xi_{2}\xi_{4}\big)+\dot w_{3}\big(\xi_{1}\xi_{2}+\xi_{3}\xi_{4}\big)\Big) \bigg\}+\nonumber\\&&
\frac{1}{R^2}\bigg\{\rho \Big[2A+\rho A'\Big]\big(\dot w_{1} ^2+\dot w_{2} ^2+\dot w_{3} ^2+2g^2-\xi_{1}\dot\xi_{1}-\xi_{2}\dot\xi_{2}-
\xi_{3}\dot\xi_{3}-\xi_{4}\dot\xi_{4}\big)-\Big[\frac{4}{\rho} A+4A'+\rho A''\Big] \cdot\nonumber\\&& \cdot \Big[\dot w_{1}\Big(w_{2}\big(\xi_{1}\xi_{2}
+\xi_{3}\xi_{4}\big)+w_{3}\big(\xi_{1}\xi_{3}-\xi_{2}\xi_{4}\big)\Big)
-\dot w_{2}\Big(w_{1}\big(\xi_{1}\xi_{2}+\xi_{3}\xi_{4}\big)-\nonumber\\&&-w_{3}\big(\xi_{1}\xi_{4}+\xi_{2}\xi_{3}\big)\Big)
-\dot w_{3}\Big(w_{1}\big(\xi_{1}\xi_{3}-\xi_{2}\xi_{4}\big)+w_{2}\big(\xi_{1}\xi_{4}+\xi_{2}\xi_{3}\big)\Big)+\nonumber\\&&
-2g^2\Big(w_{1}\big(\xi_{1}\xi_{4}+\xi_{2}\xi_{3}\big)-w_{2}\big(\xi_{1}\xi_{3}-\xi_{2}\xi_{4}\big)
+w_{3}\big(\xi_{1}\xi_{2}+\xi_{3}\xi_{4}\big)\Big)\Big]+\nonumber\\&&
2\Big[\frac{4}{\rho}A+14A'+16\rho A''+\rho^2 A'''\Big]\xi_{1}\xi_{2}\xi_{3}\xi_{4} \bigg\}+
\nonumber\\&&
\frac{1}{R^3}\bigg\{2\rho\Big[2A-\rho A'\Big]\big(w_{1} \dot w_{1}+w_{2} \dot w_{2}+w_{3} \dot w_{3}\big)g-\rho \Big[6A-6\rho A'-\rho^2 A''\Big] \cdot \nonumber\\&& \cdot \Big(\dot w_{1}\big(\xi_{1}\xi_{4}+\xi_{2}\xi_{3}\big)-\dot w_{2}\big(\xi_{1}\xi_{3}-\xi_{2}\xi_{4}\big)+
\dot w_{3}\big(\xi_{1}\xi_{2}+\xi_{3}\xi_{4}\big)\Big) \bigg\}+\nonumber\\&&
\frac{1}{R^4}\bigg\{ \rho^3\Big[2A+\rho A' \Big]g^2+\rho\Big[6A+6\rho A'+\rho^2 A'' \Big]\Big(
w_{1}\big(\xi_{1}\xi_{4}+\xi_{2}\xi_{3}\big)-w_{2}\big(\xi_{1}\xi_{3}-\xi_{2}\xi_{4}\big)+\nonumber\\&&
w_{3}\big(\xi_{1}\xi_{2}+\xi_{3}\xi_{4}\big)\Big)g+\rho \Big[ 24A+36\rho A'+12\rho^2 A''+\rho^3 A'''\Big]\xi_{1}\xi_{2}\xi_{3}\xi_{4}\bigg\}.
\eea
We also present the most general invariant action for the $(2,4,2)_{nl}$ supermultiplet induced by a global $SO(2)$ invariant prepotential. After setting
\bea\label{rho2}
\rho&=&\sqrt{z_1^2+z_2^2}
\eea
and, as before, $A(\rho)= F'(\rho)$, the lagrangian ${\cal L}_{IV}$ is expressed as
\bea\label{lagr4}
{\cal L}_{IV}&=&\Big[\frac{1}{\rho}A+A'\Big]\big(\dot z_{1}^2+\dot z_{2}^2+h_{1}^2+h_{2}^2-\eta_{1}\dot\eta_{1}-\eta_{2}\dot\eta_{2}
-\eta_{3}\dot\eta_{3}-\eta_{4}\dot\eta_{4}\big)+\frac{1}{\rho}\Big[\frac{1}{\rho^2}A-\frac{1}{\rho}A'-A'' \Big]\nonumber\\&&
\Big[\big( \dot z_{1} z_{2}-\dot z_{2} z_{1}\big)\big( \eta_{1}\eta_{2}+\eta_{3}\eta_{4}\big)-h_{1}\Big(z_{1}\big(\eta_{1}\eta_{3}-
\eta_{2}\eta_{4} \big)+z_{2}\big(\eta_{1}\eta_{4}+\eta_{2}\eta_{3}\big)\Big)-\nonumber\\&&-h_{2}\Big(z_{1}\big(\eta_{1}\eta_{4}+\eta_{2}\eta_{3} \big)-z_{2}\big(\eta_{1}\eta_{3}-\eta_{2}\eta_{4}\big)\Big) \Big]+\Big[ \frac{1}{\rho^3}A-\frac{1}{\rho^2}A'+\frac{2}{\rho}A''+A'''\Big]
\eta_{1}\eta_{2}\eta_{3}\eta_{4}+\nonumber\\&&\frac{1}{r}\bigg\{-2\Big[\frac{2}{\rho}A+A'\Big]\big(z_{1}\dot z_{1}+z_{2}\dot z_{2} \big)h_{1}-2\Big[\frac{1}{\rho}A+A'\Big]h_{2}\eta_{3}\eta_{4}-\Big[\frac{1}{\rho}A+3A'+\rho A''\Big]\cdot\nonumber\\&& \cdot\Big(\dot z_{1}\big(\eta_{1}\eta_{3}-\eta_{2}\eta_{4}\big)+\dot z_{1}\big(\eta_{1}\eta_{4}+\eta_{2}\eta_{3}\big)-h_{2}\big(\eta_{1}\eta_{2}
+\eta_{3}\eta_{4}\big)\Big) \bigg\}+\nonumber\\&&\frac{1}{r^2}\bigg\{\Big[\rho^2A'\Big]\big(\dot z_{1}^2+\dot z_{2}^2+h_{1}^2+h_{2}^2-
\eta_{1}\dot\eta_{1}-\eta_{2}\dot\eta_{2}-\eta_{3}\dot\eta_{3}-\eta_{4}\dot\eta_{4}\big)+\Big[2\rho A\Big]\big(h_{1}^2+h_{2}^2-\nonumber\\&&-
\eta_{3}\dot\eta_{3}-\eta_{4}\dot\eta_{4}\big)+\rho\Big[A+\rho A'\Big]h_{1}^2+\Big[2A'+\rho A''\Big]\big(\dot z_{1} z_{2}-\dot z_{2} z_{1}\big)\big(\eta_{1}\eta_{2}+\eta_{3}\eta_{4}\big)-\nonumber\\&&-2\Big[\frac{1}{\rho}A+A'\Big]\big(\dot z_{1} z_{2}-\dot z_{2} z_{1}\big)\eta_{3}\eta_{4}+\Big[\frac{3}{\rho}A+7A'+2\rho A''\Big]\Big(z_{1}\big(\eta_{1}\eta_{3}-\eta_{2}\eta_{4}\big)+
\nonumber\\&&z_{2}\big(\eta_{1}\eta_{4}+\eta_{2}\eta_{3}\big)\Big)h_{1}+\Big[\frac{2}{\rho}A+4A'+\rho A''\Big]\Big(z_{1}\big(\eta_{1}\eta_{4}+
\eta_{2}\eta_{3}\big)+\nonumber\\&&z_{2}\big(\eta_{1}\eta_{3}-\eta_{2}\eta_{4}\big)\Big)h_{2}+2\Big[-\frac{15}{\rho}A+14A'+12\rho A''+ \rho^2 A'''\Big]\eta_{1}\eta_{2}\eta_{3}\eta_{4}\bigg\}+\nonumber\\&&\frac{1}{r^3}\bigg\{-2\Big[\rho^2A'\Big]\big(z_{1}\dot z_{1}+z_{2}\dot z_{2}\big)h_{1}+2\rho\Big[A+\rho^2 A'\Big]h_{1}\eta_{3}\eta_{4}-\rho^2\Big[4A'-\rho A''\Big]\cdot\nonumber\\&& \cdot\Big(\dot z_{1}\big(\eta_{1}\eta_{3}-\eta_{2}\eta_{4}\big)+\dot z_{1}\big(\eta_{1}\eta_{4}+\eta_{2}\eta_{3}\big)-h_{2}\big(\eta_{1}\eta_{2}
+\eta_{3}\eta_{4}\big)\Big)\bigg\}+\nonumber\\&&\frac{1}{r^4}\bigg\{\Big[\rho^4A'\Big]h_{1}^2+\rho^2\Big[4A'+\rho A''\Big]\Big(z_{1}\big(\eta_{1}\eta_{3}
-\eta_{2}\eta_{4}\big)+z_{2}\big(\eta_{1}\eta_{4}+\eta_{2}\eta_{3}\big)\Big)h_{1}+\nonumber\\&&\rho^2\Big[4A'+8\rho A''+\rho^2 A'''\Big]\eta_{1}\eta_{2}\eta_{3}\eta_{4}\bigg\}.
\eea
Without loss of generality we could have set $R=1$ in (\ref{lagr3}) and $r=1$ in (\ref{lagr4}). It is however convenient to make explicit which terms entering the above lagrangians are due to the nonlinearity of the supertransformations. They enter (\ref{lagr3}) as powers of $\frac{1}{R}$ and
(\ref{lagr4}) as powers of $\frac{1}{r}$. The ${\cal N}=4$ invariant actions for the $(3,4,1)_{lin}$ and $(2,4,2)_{lin}$ linear supermultiplets are recovered from (\ref{lagr3}) (respectivey (\ref{lagr4})) when taking the
limit $R\rightarrow \infty$ (respectively $r\rightarrow \infty$).

\section{Prepotentials and their associated $\sigma$-models. The Uniformization.}

The ${\cal N}=4$ supersymmetric invariant actions defined in the previous Section induce $\sigma$-models $\Sigma$ which are one-dimensional
mappings on a
Riemannian target manifold ${\cal M}_g$ endowed with a metric $g$:
\bea
\Sigma&:&\begin{array}{ccl}{\bf R}&\rightarrow& {\cal M}_g,\\
 t&\mapsto &{\vec X}(t).\end{array}
\eea
${\vec X}$ denotes the local coordinates of the target manifolds. They correspond to the physical bosonic component fields
(called, for this reason, ``target coordinates") entering the off-shell supermultiplets. The remaining bosonic components are the auxiliary fields. \par
The associated $\sigma$-models are constructed by\\
{\em i}) consistently setting equal to zero all the fermionic fields in the supermultiplets,\\
{\em ii}) solving the algebraic equations of motion for the auxiliary fields,\\
{\em iii}) reexpressing the resulting lagrangians as ${\cal L} = g_{ij} {\dot X}^i{\dot X}^j$. \par
The metric $g_{ij}$ is a functional of the prepotential $F({\vec X})$, s.t. $\relax g_{ij}\equiv g_{ij}[F({\vec X})]$.\par
For the $(3,4,1)_{nl}$ supermultiplet with prepotential $F(\rho)$ (\ref{lagr3}) the metric is diagonalized when expressed in terms of the redefined
target coordinates $\rho, \theta_1,\theta_2$ s.t.
\bea
&w_{1}=\rho cos(\theta_{1})sin(\theta_{2}),\quad
w_{2}=\rho sin(\theta_{1})sin(\theta_{2}),\quad
w_{3}=\rho cos(\theta_{2}).&
\eea
The non-vanishing components of the metric ($g_{\rho\theta_{1}}=g_{\rho\theta_{2}}=g_{\theta_{1}\theta_{2}}=0$) are
\bea\label{metr341nl}
g_{\rho\rho}&=&\frac{4\big(\rho^2+1\big)}{\rho}\Big[\rho F''(\rho)+F'(\rho)\Big],
\nonumber\\
g_{\theta_{1}\theta_{1}}&=&\rho\big(\rho^2+1\big)sin(\theta_{2})\Big[\rho F''(\rho)+F'(\rho)\Big],
\nonumber\\
g_{\theta_{2}\theta_{2}}&=&\rho\big(\rho^2+1\big)\Big[\rho F''(\rho)+F'(\rho)\Big].
\eea
For the $(2,4,2)_{nl}$ supermultiplet with prepotential $F(\rho)$ (\ref{lagr4}) the metric is diagonalized in terms of the redefined target coordinates $\rho, \alpha$ s.t.
\bea
&z_1=\rho \cos (\alpha),\quad z_2=\rho\sin(\alpha).&
\eea
The non-vanishing components of the metric ($g_{\rho\alpha}=0$) are
\bea\label{metr242nl}
g_{\rho\rho}&=&\bigg\{\rho^2\Big[F''(\rho)\Big]^2\big(4\rho^6+9\rho^4+6\rho^2+1\big)+2\rho\Big[F''(\rho)\Big]\Big[F'(\rho)\Big]\big(5\rho^4
+6\rho^2+1\big)+\nonumber\\&&\Big[F'(\rho)\Big]^2\big(6\rho^2+1\big)\bigg\}\Bigg/\rho\bigg\{\rho\Big[F''(\rho)\Big]\big(\rho^2+1\big)^2+
\Big[F'(\rho)\Big]\big(3\rho^2+1\big)\bigg\},\nonumber\\
g_{\alpha\alpha}&=&\rho\bigg\{\Big[F'(\rho)\Big]+\rho\big(\rho^2+1\big)\Big[F''(\rho)\Big]\bigg\}.
\eea
An inverse problem can be defined. It consists in the determination (from (\ref{metr341nl}) or (\ref{metr242nl})) of a prepotential $F$ which reproduces
a given reference metric ${\widehat g}_{ij}$.
\par
For linear supermultiplets (see (\ref{lagr1}) and (\ref{lagr2})) the induced metric $g_{ij}$ is conformally flat ($g_{ij}=\Phi({\vec X}\delta_{ij}$).
In particular the constant flat metric is recovered from a quadratic prepotential. For a two-dimensional target manifold a special case of the inverse problem is the century-old Uniformization Problem discussed by Liouville. A conformally flat metric admits constant curvature everywhere if the (suitably normalized) conformal factor $\Phi$ satisfies the Liouville's equation $\Box \Phi=\exp(\Phi)$.\par
For the non-linear supermultiplets the choice of a quadratic prepotential ($F\propto \rho^2$) does not reproduce a constant metric. The curvature tensor
${\cal R}_{ijkl}$, the Ricci tensor ${\cal R}_{ij}$ and the curvature scalar ${\cal R}$ can be computed from (\ref{metr341nl}) and (\ref{metr242nl}). \par
It is useful to present some results for the $(2,4,2)_{nl}$ supermultiplet
(we will follow the conventions of (\cite{wei}) which, in particular, sets ${\cal R}=-2$ for a sphere of radius $r=1$).  The curvature ${\cal R}$ associated to the quadratic prepotential
\bea
F&=& C\rho^2
\eea
is given by
\bea\label{curvsc}
{\cal R}&=&\big[\rho^2-44\big]\big/ \big[ C(\rho^2+1)^2(\rho^2-8)^2 \big].
\eea
Setting $C=\frac{11}{32}$ normalizes ${\cal R}(0)=-2$ at the origin.\par
In this context the Uniformization Problem can be attacked through Taylor-expansion as follows. One can express $F$ in powers of $\rho^2$
($F=\sum_{i=1} C_i {\rho^{2i}}$) and adjust the coefficients in order to make vanishing the $\rho$-derivatives of ${\cal R}$ at the origin.
If we set, e.g.,
\bea
F&=& N(\rho^2+k\rho^4)
\eea
(that is $C_1=N$, $C_2=kN$, $C_j=0$ for $j=2,3,\ldots$) we obtain, for the corresponding curvature scalar ${\cal R}$,
\bea\label{curvsc2}
{\cal R}(0)&=& \big[(8k+1)(8k^2+18k+1)\big]\big/\big[ 8(2k+1)(2+k)N\big],\nonumber\\
{\cal R}'' |_{\rho=0}&=& -\frac{1536k^7+6400k^6+8744k^5+4332k^4+266k^3-227k^2-39k+12}{4k(2k+1)^2(2+k)^2N}
\eea
(the odd-derivatives of ${\cal R}$ are all vanishing at $\rho=0$).\par
Requiring ${\cal R}''|_{\rho=o}=0$ implies that $k$ is fixed to be a root of the $7^{th}$-order polynomial at the numerator of the r.h.s. of the second equation. Its unique real root is
\bea
k&\approx& -1.97997.
\eea
The normalization ${\cal R}(0)=-2$ gives for $N$ the approximate numerical solution
\bea
N&\approx& 51.2779.
\eea
This choice of the prepotential produces a metric whose curvature is approximately constant in the neighborhood of the origin.

\section{Some remarks on the supersymmetric extension of the $2^{nd}$ and $3^{rd}$ Hopf maps.}

It is worth to point out some basic features of the supersymmetric extensions of the second and third Hopf maps (that is, the supersymmetric extensions of the (\ref{hopfd1}) diagram for $k=4,8$). In \cite{gknty}
the supersymmetric extension of the bilinear mapping $p: {\bf R}^8\rightarrow {\bf R}^5$ was introduced and applied to the construction of supersymmetric systems in the presence of an $SU(2)$ Yang monopole. The supersymmetric extension of the Hopf map (the entire (\ref{hopfd1}) diagram) would require, of course, the restrictions on the spheres.
We will briefly summarize the results of \cite{gknty} (see also \cite{bty}) and make further comments.\par
The $(8,8)$ root supermultiplet corresponds to a minimal, linear, off-shell representation for all values  ${\cal N}=5,6,7,8$. On the other hand the $p$ map is (globally) $SU(2)$-invariant. In accordance with the Schur's lemma, whose supersymmetric extension is presented in Section {\bf 9}, ${\cal N}=5$ is the maximal number of supersymmetric generators acting on $(8,8)$ and commuting with the $su(2)$ algebra. The $k=4$ supersymmetric extension of $p$ produces the ${\cal N}=5$ linear mapping $(8,8)\rightarrow (5,11,10,5,1)$.
The latter supermultiplet is a linear dressing of the ${\cal N}=5$ non-minimal ``enveloping'' (see \cite{gknty}) supermultiplet whose field content is given by the (${\cal N}=5$) Newton's binomial (the linear dressing maps $(1,5,10,10,5,1)\mapsto(0,5,11,10,5,1)$). Four of the five supercharges can be picked up to
construct, with the method illustrated in Section {\bf 5}, a manifestly ${\cal N}=4$ off-shell invariant action depending on an unconstrained prepotential depending on five bosonic target coordinates. Imposing the invariance under the fifth supercharge puts a constraint on the superpotential.\par
The compatibility of this construction with the restriction on spheres was not investigated in \cite{gknty}.
This program can now be completed by applying the methods discussed in the present paper. The use of the stereographic projection allows us to produce non-linear off-shell realizations for ${\cal N}>4$. It is expected that the restriction ${\bf R}^8\rightarrow {\bf S}^7$ would produce an ${\cal N}=8$ non-linear
realization of supersymmetry which is local and could be expressed in terms of the octonionic structure constants. The use of the hyperspherical coordinates on the other hand would induce non-local realizations
of the supersymmetry (based on ratio of trigonometric functions) which, on the other hand, are covariant under
the $SU(2)$ group of transformations. These extensions are currently under investigation.\par
There is no obstruction in further applying these methods to the third Hopf map. The bosonic bilinear map $p$ sends now
${\bf R}^{16}\rightarrow {\bf R}^9$. A $(16,16)$ root supermultiplet carries a minimal linear representation for ${\cal N}=9$. The supersymmetric extension of the map $p$ sends the root supermultiplet into the
non-minimal ${\cal N}=9$ linear supermultiplet of field content $(9, 37, 84,126,126,84,36,9,1)$, which is a linear dressing
of the ${\cal N}=9$ ``enveloping supermultiplet'' based on the ${\cal N}=9$ Newton's binomial
($(1,9,36, 84, 126, 126, 84,36,9,1)\mapsto (9, 37, 84,126,126,84,36,9,1)$).  As before, the restrictions on spheres can be presented in terms of the stereographic projection or the hyperspherical parametrization.

\section{Some remarks on the oxidation program.}

A major area of applications of the construction and classification of one-dimensional ${\cal N}$-extended off-shell
realizations (linear and minimal, linear and non-minimal, non-linear) of supersymmetry and their invariant actions concerns the so-called ``oxidation program'' \cite{glpr,top}, i.e. the reconstruction of higher-dimensional supersymmetric theories from
$1D$ supersymmetric data. Several types of supermultiplets of given field content arises from the dimensional reduction of known supersymmetric theories. For instance, the reduction to a $0+1$ quantum mechanical system
of the ${\cal N}=4$ Super-Yang-Mills theory in $3+1$ dimension produces a supermultiplet of field content $(9,16,7)$ which carries an off-shell representation of $9$ supercharges. The remaining $7$ supersymmetry generators ($=16-9$) can only be realized on-shell \cite{top} (this is in consequence of the \cite{pt} result and counting). Similarly, for the maximal, eleven-dimensional supergravity, the transverse-coordinate supermultiplet containing the graviton, the gravitinos and the $3$-form admits field content $(44,128,84)$. As such, it carries an off-shell representation for ${\cal N}=16$ supercharges. The remaining $16$ supersymmetry generators which complete the total number of $32=8\times 4$ supercharges only close
on-shell. Due to the \cite{pt} counting, an off-shell formulation of the eleven-dimensional M-theory would require (if it exists) a supermultiplet with at least $32,768=2^{15}$ bosonic component fields and an equal number of fermionic component fields. These features have been discussed in greater detail in \cite{top} and will not be further repeated here. In this paper it is sufficient to mention that the oxidation program can be carried out in two steps. In the first step an ${\cal N}$-extended, one-dimensional, supersymmetric theory has to be constructed for a sufficiently large value of ${\cal N}$. The most interesting values for ${\cal N}$ correspond to ${\cal N}=4,8,16,32$ which can be associated to a dimensional reduction of a $4$, $6$, $10$, or $11$-dimensional supersymmetric theory, respectively. This first step produces a one-dimensional supersymmetric theory possessing a necessary, but not sufficient condition allowing its oxidation to a higher-dimensional supersymmetric theory. The second step consists in deriving the conditions on the $1D$ supersymmetric data in order to reconstruct the oxidized data. For instance, questions to be answered concern the organization of the $1D$ supermultiplets in representations of the higher-dimensional Lorentz groups. This second part of the program, see the references \cite{fil,fl}, is still at its infancy.

Further comments on the possible relevance of the present results to the oxidation program will be made in
the Conclusions.

\section{The Schur's lemma extended to minimal, linear, supermultiplets.}

The Schur's lemma \cite{oku} is a statement about the most general matrix $S$ commuting with all the $p+q$ gamma matrices $\gamma_i$ ($i=1,\ldots, p+q$) which define the $C(p,q)$ Clifford algebra over the reals
($p$ matrices of the set have square $+{\bf 1}$ and $q$ matrices have square $-{\bf 1}$). Depending on the $(p,q)$ pair, the three following
cases are obtained:
\\
{\em i}) the real case (${\bf R}$) s.t. $S= \lambda_0{\bf 1}$, \\
{\em ii}) the almost complex case (${\bf C}$) s.t. $S$ is given by the sum $S=\lambda_0{\bf 1} + \lambda_1{\tau_1}$, with ${\tau_1}^2=-{\bf 1}$,\\
{\em iii}) the quaternionic case (${\bf H}$) s.t. $S$ is given by the sum $S= \lambda_0 {\bf 1} + \sum_{j=1}^3\lambda_j\tau_j$, \par with
$[\tau_j,\tau_k]=\epsilon_{jkl}\tau_l$ and ${\tau_j}^2=-{\bf 1}$.\\
In the above formulas $\lambda_k$'s are real numbers.\par
The one-to-one connection, pointed out in \cite{pt}, between irreducible representations of Clifford algebras and representation of the ${\cal N}$-extended superalgebra (\ref{superalgebra}) given by the minimal linear supermultiplets
of field content $(n,n)$ (the ``root"' supermultiplets) implies that the most general matrix $S$ commuting with the
${\widetilde{Q}}_I$ ($I=1,\ldots, {\cal N}$) supersymmetry operators of the root representation is directly read from the associated Clifford irrep.\\
The higher length minimal linear supermultiplets (see \cite{krt}) of field content $(n_1,n_2,n_3,\ldots)$ are expressed in terms of the supersymmetry operators
${\widehat Q}_I= D{\widetilde Q}_I D^{-1}$, where $D$ is the diagonal dressing operator whose entries are $1$ and powers of the time-derivative operator $\partial_t$. The most general $S$ commuting with the dressed operators
${\widehat Q}_I$ is recovered by imposing the further condition $[S,D]=0$. As a consequence, a necessary but not sufficient condition for $S$ to be of ${\bf C}$ type (${\bf H}$ type) is that the set of integers $n_i$ entering the field
content are all even numbers (all multiples of 4).  \par
We are now in the position to write down the Schur's type (${\bf R}$, ${\bf C}$, ${\bf H}$) of all inequivalent minimal linear supermultiplets up to ${\cal N}\leq 8$. For ${\cal N}\neq 5,6$, such inequivalent supermultiplets are uniquely
characterized by their field content. Their complete list is given in \cite{krt}. For ${\cal N}=5,6$ inequivalent linear supermultiplets with the same field content, but differing in connectivity (of the associated graph, see \cite{kt}) are encountered. Their admissible connectivities, expressed through the $\psi_g$ symbol, are classified in \cite{kt} (see also \cite{kt2}).\par
The following results are obtained:\\
- for ${\cal N}=1,7,8$ all minimal linear supermultiplets are of ${\bf R}$ type,\\
- for ${\cal N}=2$ the $(2,2)$ supermultiplet is of ${\bf C}$ type ($(1,2,1)$ is of ${\bf R}$ type),\\
- for both ${\cal N}=3,4$ the $(4,4)$ supermultiplet is of ${\bf H}$ type, $(2,4,2)$ is of ${\bf C}$ type, while the remaining supermultiplets are of ${\bf R}$ type. \par
For ${\cal N}=5$ the results are summarized in the following table. The Schur's type is reprorted in the last column. The $\psi_g$ connectivity of reference \cite{kt} is reported in the third column. The decomposition into ${\cal N}=4$ supermultiplets (see \cite{kt2}) is reported in the second column. The labels ($A,B,C$) are introduced to distinguish supermultiplets with the same field content.
 \begin{eqnarray}
\begin{tabular}{|l|l|l|c|c|}
\hline
fields cont. & ${\cal N}=4$ decomp. & $\psi_g$ connectivities  & labels & Schur's type\\
\hline\hline
$(8,8)$  & $(4,4)+(4,4)$ & $8_0$ &&${\bf H}$\\
\hline\hline
$(1,8,7)$  & $(0,4,4)+(1,4,3)$ & $3_5+5_4$ &&${\bf R}$\\
\hline
$(2,8,6)$  & $(0,4,4)+(2,4,2)$ & $2_5+2_4+4_3$&$A$ &${\bf C}$\\
  & $(1,4,3)+(1,4,3)$ & $6_4+2_3$ &$B$ &${\bf R}$\\
\hline
$(3,8,5)$  & $(0,4,4)+(3,8,5)$ & $1_5+3_4+4_2$ &$A$&${\bf R}$\\
  & $(1,4,3)+(2,4,2)$ & $2_4+5_3+1_2$ &$B$&${\bf R}$\\
\hline
$(4,8,4)$  & $(0,4,4)+(4,4,0)$ & $4_4+4_1$ &$A$&${\bf H}$\\
  & $(1,4,3)+(3,4,1)$ & $1_4+3_3+3_2+1_1$ &$B$&${\bf R}$\\
   & $(2,4,2)+(2,4,2)$ & $4_3+4_2$ &$C$&${\bf C}$\\
   \hline
   $(5,8,3)$  & $(1,4,3)+(4,4,0)$ & $4_3+3_1+1_0$ &$A$&${\bf R}$\\
  & $(2,4,2)+(3,4,1)$ & $1_3+5_2+2_1$&$B$&${\bf R}$ \\
\hline
$(6,8,2)$  & $(2,4,2)+(4,4,0)$ & $4_2+2_1+2_0$&$A$&${\bf C}$ \\
  & $(3,4,1)+(3,4,1)$ & $2_2+6_1$ &$B$&${\bf R}$\\
\hline
$(7,8,1)$  & $(3,4,1)+(4,4,0)$ & $5_1+3_0$&&${\bf R}$ \\
\hline\hline
$(1,5,7,3)$&
$(1,4,3)+(0,1,4,3)$&$5_4$&&${\bf R}$\\
$(1,6,7,2)$  & $(1,4,3)+(0,2,4,2)$ &$1_5+5_4$ &&${\bf R}$ \\
$(1,7,7,1)$  & $(1,4,3)+(0,3,4,1)$ & $2_5+5_4$&&${\bf R}$ \\
$(2,6,6,2)$  & $(2,4,2)+(0,2,4,2)$ & $2_4+4_3$&&${\bf C}$ \\
$(2,7,6,1)$  & $(2,4,2)+(0,3,4,1)$ & $1_5+2_4+4_3$&&${\bf R}$ \\
$(3,7,5,1)$ &  $(3,4,1)+(0,3,4,1)$ & $3_4+4_2$ &&${\bf R}$\\
\hline
\end{tabular}
\end{eqnarray}
A similar table is produced for the ${\cal N}=6$ minimal linear supermultiplets. It is given by
 \begin{eqnarray}
\begin{tabular}{|l|l|c|c|}
\hline
fields cont. &  $\psi_g$ connectivities  & labels & Schur's type\\
\hline\hline
$(8,8)$  &  $8_0$ &&${\bf C}$\\
\hline\hline
$(1,8,7)$  &  $2_6+6_5$ &&${\bf R}$\\
\hline
$(2,8,6)$  & $2_6+6_4$&$A$ &${\bf C}$\\
  &  $4_5+4_4$ &$B$ &${\bf R}$\\
\hline
$(3,8,5)$  & $2_5+2_4+4_3$ &$A$&${\bf R}$\\
  &  $6_4+2_3$ &$B$&${\bf R}$\\
\hline
$(4,8,4)$  &  $4_4+4_2$ &$A$&${\bf C}$\\
  &  $2_4+4_3+2_2$ &$B$&${\bf R}$\\
   &  $8_3$ &$C$&${\bf R}$\\
   \hline
   $(5,8,3)$  &  $4_3+2_2+2_1$ &$A$&${\bf R}$\\
  &  $2_3+6_2$&$B$&${\bf R}$ \\
\hline
$(6,8,2)$  &  $6_2+2_0$&$A$&${\bf C}$ \\
  &  $4_2+4_1$ &$B$&${\bf R}$\\
\hline
$(7,8,1)$  & $6_1+2_0$&&${\bf R}$ \\
\hline\hline
$(1,6,7,2)$  & $6_5$ &&${\bf R}$ \\
$(1,7,7,1)$  &  $1_6+6_5$&&${\bf R}$ \\
$(2,6,6,2)$  &  $6_4$&&${\bf C}$ \\
$(2,7,6,1)$  &  $1_6+6_4$&&${\bf R}$ \\
\hline
\end{tabular}
\end{eqnarray}

\section{Conclusions.}

Let us first summarize the main results of the present paper.\par
We explicitly constructed the supersymmetric extension of the first Hopf map, which results in connecting four ${\cal N}=4$ one-dimensional off-shell
supermultiplets. As a consequence, the linear $(3,4,1)$ supermultiplet, as well as the non-linear $(3,4,1)_{nl}$ and $(2,4,2)_{nl}$
supermultiplets, are induced from the ${\cal N}=4$ $(4,4)$ ``root" linear supermultiplet, whose four bosonic target coordinates can be regarded as a parametrization of ${\bf R}^4$. The stereographic projection acting on the bosonic target coordinates is used to induce the non-linear supermultiplets.
Such a non-linearity is {\em local}, i.e., the supertransformations are differential polynomials in the component fields entering the supermultiplets.
We further applied a construction introduced in previous works \cite{krt, grt} to obtain the one-dimensional ${\cal N}=4$ off-shell invariant actions associated to each supermultiplet. Each such an action depends on an unconstrained prepotential of its bosonic target coordinates. An Inverse Problem for the prepotential corresponds to its determination in order to reproduce (in the purely bosonic sector of the theory) a $\sigma$-model of a given target metric ${\widehat g}_{ij}$ of reference.\par
The same procedure can be rather straightforwardly applied to the supersymmetric extensions of the second and third Hopf maps. The main differences with respect to the first Hopf map case have been outlined. The first and the second Hopf maps admit ${\bf S}^1$ (respectively ${\bf S}^3$) as a fibration. The covariance under the $U(1)$ or the $SU(2)$ group would require the use of another parametrization (for instance the hyperspherical coordinates) instead of the stereographic projection to induce the non-linear supermultiplets. There is a price to be paid in this case, however, since the resulting non-linear supertransformations are {\em non-local}. The construction of the non-linear supertransformations under hyperspherical coordinates is under investigation and left for future works. In the bosonic case the existence of the $U(1)$ or the $SU(2)$ fiber can be regarded to be a consequence of the Schur's lemma. For completeness, we extended here the Schur's lemma to all  minimal, linear, off-shell supermutiplets up to ${\cal N}\leq 8$, listing the ones which commute with the $u(1)$ or the $su(2)$ algebra generators.\par
The theory of non-linear off-shell realizations of one-dimensional ${\cal N}$-extended supersymmetry is rather poorly understood. It is based on a set of consistency conditions to be fulfilled. Finding their general solution for a given, generic, value of ${\cal N}$ is a formidable task. The determination of their inequivalent classes (under reparametrization of their component fields) is another extremely hard task (both these problems are currently investigated with brute-force techniques). The {\em local} non-linear supertransformations are more manageable. For instance, one is guaranteed that the local, non-linear supermultiplets here obtained are not equivalent to the linear supermultiplets with the same field content. The $(3,4,1)_{nl}$ and the $(2,4,2)_{nl}$ non-linear supermultiplets have a very compelling geometrical origin and are nicely formulated (the fact, e.g., that $(3,4,1)_{nl}$ can be expressed through the quaternionic structure constants). We leave for future works the analysis of their possible relations with analogous non-linear supermultiplets of the same field content, previously obtained in the literature in terms of different constructions, see \cite{root, bny}.\par
Concerning the possible applications of the present results (as well as their further extensions to the second Hopf map) we can mention the investigation concerning the motion of superparticles in the presence of an $U(1)$ or a $SU(2)$ Yang monopole (extending the analysis of \cite{gknty} to a non-flat geometry).\par
Let's say some final words concerning the applications to the oxidation program. We have already commented, in this context, about the importance of the construction of the most general class of linear and non-linear one-dimensional off-shell realizations of supersymmetry. We limit here just to mention that a very promising line of investigation concerns the one-dimensional Supersymmetric Quantum Mechanics viewpoint concerning the topological twist of SuperYang-Mills Theory \cite{bbbm}, as well as the dimensional reduction of the maximal supergravity (associated with billiards, with a special role played by the $E_{10}$ algebra \cite{e10,bill}).  \par

{}~
\\{}~
\par {\large{\bf Acknowledgments}}{} ~\\{}~\par

We are grateful to S. Krivonos and A. Nersessian for useful discussions.\\
This work has been supported by Edital Universal CNPq Proc. 472903/2008-0.

{}~
\\{}~
\par {\Large{\bf Appendix:
quaternionic covariance of the ${\cal N}=4$ $(3,4,1)_{nl}$ non-linear supermultiplet.}}{} ~\\{}~\par

The non-linear ${\cal N}=4$ $(3,4,1)_{nl}$ supermultiplet (\ref{table3}) can be expressed in terms of the
$su(2)$ (or quaternionic) structure constants $\delta_{ij}$ and $\epsilon_{ijk}$. At most one of the four supersymmetry generators can be linearly realized (it will be denoted as ``$Q_4$''). The three remaining
supersymmetry generators $Q_i$ ($i=1,2,3$) are non-linearly realized. The covariant basis for $(3,4,1)_{nl}$ is written in terms of the bosonic fields ${\widehat w}_i, {\widehat g}$ and the fermionic fields ${\widehat \xi}_i, {\widehat \xi}$. Their explicit expression in terms of the component fields entering the $(4,4)$ root multiplet (\ref{table1}) is given by (without loss of generality we have set $R=1$, see (\ref{table3}))
\bea
&\begin{array}{ll} {\widehat w}_i = \frac{u_i}{1-u_4},&g=\frac{{\dot u}_4}{1-u_4},\\
{\widehat \xi}_i = \frac{\psi_i}{1-u_4}+\frac{u_i\psi_4}{(1-u_4)^2}, & {\widehat \xi}=\frac{\psi_4}{1-u_4}.
\end{array}
&
\eea
The supersymmetry transformations, acting on ${\widehat w}_i, {\widehat \xi}_i, {\widehat \xi}, {\widehat g}$,
are given by
\bea
&\begin{array}{ll} Q_4 {\widehat w}_i = {\widehat \xi} ,& Q_4 {\widehat g}={\dot {\widehat \xi}}_4,\\
Q_4{\widehat \xi}_i = {\dot {\widehat w}}_i, & Q_4{\widehat \xi}={\widehat g},
\end{array}
&
\eea
together with
\bea
Q_i{\widehat w}_j&=& \epsilon_{ijk} ({\widehat \xi}_k -{\widehat w}_k{\widehat \xi}) -(\delta_{ij}+{\widehat w}_i{\widehat w}_j){\widehat \xi}+ {\widehat w}_j{\widehat \xi}_i,\nonumber\\
Q_i{\widehat \xi}_j&=& -\epsilon_{ijk} ({\dot{\widehat w}_k}-{\widehat w}_k{\widehat g} -{\widehat \xi}_k{\widehat \xi})-{\widehat w}_j({\dot{\widehat w}}_i-{\widehat w}_i{\widehat g})+({\widehat w}_i{\widehat \xi}_j+{\widehat w}_j{\widehat \xi}_i){\widehat \xi} +\delta_{ij}{\widehat g},\nonumber\\
Q_i{\widehat \xi}&=&-{\widehat \xi}{\widehat \xi}_i-{\dot {\widehat w}}_i g ,\nonumber\\
Q_i {\widehat g} &=& {\dot{\widehat \xi}}_i-{\dot {\widehat w}_i}{\widehat \xi}+{\widehat w}_i{\dot{\widehat \xi}}.
\eea
One should note that in this covariant basis for $(3,4,1)_{nl}$ with one linearly realized supersymmetry generator, the right hand side is no longer bilinear in the fields since $Q_i{\widehat \xi}_j$ contains a trilinear term.

\end{document}